\documentclass[prl,showpacs,floatfix,twocolumn,amsmath,amssymb,dvips]{revtex4-1}

\usepackage{epsfig}
\usepackage{graphicx}
\usepackage{dcolumn}
\usepackage{bm}
\usepackage[colorlinks=true,dvipdfm]{hyperref}

\begin{document}
\title{Is an imaginary fixed point physical or unphysical?}

\author{Fan Zhong}
\affiliation{State Key Laboratory of Optoelectronic Materials and
Technologies, School of Physics and Engineering, Sun Yat-sen
University, Guangzhou 510275, People's Republic of China}

\date{\today}

\begin{abstract}
It has been proposed that a first-order phase transition driven to happen in the metastable region exhibits scaling and universality near an instability point controlled by an instability fixed point of a $\varphi^3$ theory. However, this fixed point has an imaginary value and the renormalization-group flow of the $\varphi^3$ coupling diverges at a finite scale. Here combining a momentum-space RG analysis and a nucleation theory near the spinodal point, we show that imaginary rather than real values are physical counter-intuitively and thus the imaginary fixed point does control the scaling.
\end{abstract}

\pacs{05.70.Fh, 64.60.My, 64.60.ae, 64.60.Q-}
\maketitle

Nowadays, phase transitions are usually
classified into continuous phase transitions and discontinuous or
first-order phase transitions (FOPT) \cite{fisherc}. Critical phenomena, with the hallmark of
scaling and universality \cite{Stanley,Barmatz} exhibited near a continuous phase transition, have been well accounted for by the
renormalization-group (RG) theory \cite{Wilson,Ma}. For a generic FOPT driven beyond its equilibrium transition point by an external field in a usual $\phi^4$ theory below its critical point, it has been suggested that a field-theoretical RG theory of a derived $\varphi^3$ model can also well account for the universal scaling behavior found near such a driven FOPT \cite{zhongl05,zhong11}. A salient feature is that here the fixed point of the $\varphi^3$ coupling is imaginary in value. To reach such an imaginary fixed point, an initial coupling with a small imaginary part is sufficient \cite{zhong11}, which has been confirmed in a nonperturbative RG analysis \cite{li12}. However, for a physical coupling of purely real initial value, the RG flow has to diverge at a finite scale in order to achieve an imaginary part. This leads to concern as to whether the perturbation analysis is still valid and thus whether such a theory can indeed describe the scaling behavior of the FOPT or not.

To answer these questions, we need to know what physically this finite scale is, why it leads to divergence, and where the imaginary part comes from. Here, combining a momentum-shell integration RG analysis and a nucleation theory \cite{Langer67} near the spinodal point \cite{Klein83,Unger84}, we show that, at the finite scale, the free-energy cost for nucleation out of the metastable state in which the system lies and the metastable well itself vanish. This places the system exactly at a true instability point and exhibits the divergence. The integration for the partition function then diverges and has to be analytically continued to complex plane in order to be physically meaningful. As a result, the system enters the imaginary domain and can thus reach the imaginary fixed point. Therefore, counter the intuition that only real values are physical, here imaginary values are physical instead. The divergence at the characteristic scale just signals the irrelevant degrees of freedom have been eliminated and the true instability point has been reached and thus does not spoil the perturbation RG analysis.

First we briefly derive the $\varphi^3$ model. Consider the usual $\phi^4$ model with a Ginzburg-Landau Hamiltonian
\begin{equation}
{\cal H}_4[\phi ] = \int d{\bf x}\left[ \frac{1}{2}r\phi ^2
+ \frac{1}{4!}g\phi ^4 + \frac{1}{2}K(\nabla \phi)^2-H\phi \right]
\label{h4}
\end{equation}
for a scalar order parameter $\phi$ in the presence of its conjugate field $H$, where $ r$ is the reduced temperature and $g$ a coupling constant. The constant $K$ has also been introduced for the later RG study. Although metastability is a dynamic process, statics is sufficient for our present purpose.
In the mean-field approximation, it is well known that ${\cal H}_4$ has a critical point at $r=0$ and $H=0$. On the other hand, for each $r<0$, there is an FOPT between two equilibrium phases, each with an order parameter of opposite sign to the other. However, this transition can only take place beyond the spinodal point at which the barrier between the two phases vanishes in the same approximation. In order to study the property of the transition, we shift the order parameter by $\phi=M_0+\varphi$ with a uniform $M_0$, neglect the irrelevant terms, and arrive at the $\varphi^3$ model \cite{zhongl05,zhong11}
\begin{equation}
{\cal H}[\varphi] = \int d{\bf x} \left[
\frac{1}{2}\tau\varphi^2 +\frac{1}{3!}v\varphi^3+
\frac{1}{2}K(\nabla \varphi)^2 -h\varphi \right], \label{h3}
\end{equation}
which has the spinodal at $\tau=r+gM_0^2/2=0$ and $h=H-rM_0-gM_0^2/3!=0$ as its mean-field instability point, where $v=gM_0$. Given ${\cal H}$, the partition function is obtained by a functional integration over all configurations
\begin{equation}
Z=\int {\cal D}\varphi\exp{(-{\cal H})}\label{z}
\end{equation}
and the free energy is $F=-\ln Z$. In the following, we shall concentrate on the model at $h=0$. In fact, a shift of $\varphi$ can change $h$ to $\tau$ and vice versa and only one of them is independent.

\begin{figure}[t]
\centerline{\epsfig{file= 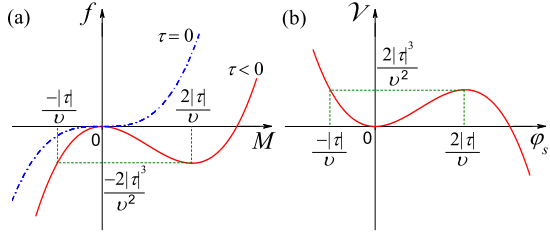,width=\columnwidth}}
\vskip -2.5cm
\caption{\label{foptfr}(Color online) (a) $\varphi^3$ mean-field free-energy density $f$ for different $\tau$. (b) The potential ${\cal V}=-f$ for $\tau<0$.}
\end{figure}

In the mean-field approximation in which $\varphi$ assumes a uniform most probable configuration $M$, the free energy density is
\begin{equation}
f(M)= F/V=\frac{1}{2}\tau M^2 + \frac{1}{3!}gM^3, \label{f}
\end{equation}
where $V$ is the system volume. As seen in Fig.~\ref{foptfr}(a), for $\tau\neq0$, there is a metastable minimum and a saddle point at $M_m=-2\tau/v$ and $M_s=0$, respectively, for $\tau<0$ with a difference in free-energy density $\Delta f=2|\tau|^3/3v^2$. When $\tau>0$ similar behavior appears; only their roles exchange. So, we shall assume $\tau<0$ for definiteness as we shall see below that its fixed point value is of this sign. The transition from $M_m$ and $M_s$ happens at $\tau=0$, the spinodal point, at which both states merge. This corresponds to the real system moving from its own metastable to the spinodal point, at which the real FOPT takes place corresponding in turn to the transition from $M_s$ to the unbounded left side in Fig.~\ref{foptfr}(a). Although the two transitions appear different, they occur at the same spinodal and it can be shown that in its vicinity the mean-field exponents describe correctly the real FOPT \cite{zhong11}. This supports strongly the relevancy of the $\varphi^3$ theory.

Next we present the momentum-space RG study of the model, which also serves as an introduction to the problem. Comparing ${\cal H}$ with ${\cal H}_4$, one sees that the role played by the spinodal point to the $\varphi^3$ model is similar to that played by the critical point to the $\phi^4$ theory. This prompts an RG study to take into account fluctuations. The usual RG consists in first integrating out the order parameter with wave numbers between the cutoff $\Lambda$ and $\Lambda/s$ with a scaling factor $s\geq1$ and then rescaling the length to restore the cutoff. The remaining order parameter has also to be rescaled accordingly in order to ensure invariant \cite{Wilson,Ma}. A previous RG study of a model with both $\phi^4$ and $\phi^3$ couplings has been performed to investigate pseudocritical phenomena near the spinodal point, with a conclusion that a true spinodal does not exist \cite{Saito}. Collecting those terms relevant to ${\cal H}$ and setting $s=1+\delta l$, we find the differential RG equations \cite{Saito}
\begin{subequations}
\label{rg}
\begin{eqnarray}
\frac{\partial K}{\partial l}&=&-\eta K-\frac{4-d}{2d}A_6v^2K,\label{k}\\
\frac{\partial v}{\partial l}&=&\left(\frac{1}{2}\epsilon-\frac{3}{2}\eta\right)v+A_6v^3,\label{v}\\
\frac{\partial \tau}{\partial l}&=&(2-\eta)\tau-\frac{1}{2}A_4v^2,\label{r}
\end{eqnarray}
\end{subequations}
where $\eta$ is an exponent, $d$ the space dimensionality, $\epsilon=6-d$, and $A_n=K_d\Lambda^{d-6}K^{n/2}$ with $K_d=2/[(4\pi)^{d/2}\Gamma(d/2)]$ ($\Gamma$ is the gamma function). The flow equation~(\ref{rg}) has, besides the Gaussian fixed point at $\tau=0$ and $v=0$, a nontrivial fixed point
\begin{subequations}
\label{rgs}
\begin{eqnarray}
v^{*2}&=&-\frac{2d}{(12+d)A_6}\epsilon\equiv-\frac{A}{A_6}\epsilon,\label{vs}\\
\eta^*&=&-\frac{4-d}{2d}A_6v^{*2}=\frac{4-d}{12+d}\epsilon,\label{es}\\
\tau^*&=&\frac{1}{4}A_4v^{*2}=-\frac{AA_4}{4A_6}\epsilon,\label{rs}
\end{eqnarray}
\end{subequations}
which is, indeed, the ``instability'' fixed point with $u^{*2}=v^{*2}A_6=-2\epsilon/3$ and the ``instability'' exponent $\eta=-\epsilon/9$ to one-loop order found by a field-theoretical RG method \cite{zhongl05,zhong11} when $d=6$, the dimension on which the $\epsilon$ expansion is based. Although the fixed point is imaginary in value, one can show from Eqs.~(\ref{rg}) and (\ref{rgs}) that it is infrared stable for $d<6$ similar to the fixed point for critical phenomena and thus controls the large-scale behavior. There exist also evidences that this fixed point influences the flows of effective exponents in a Monte Carlo RG study of the FOPTs in the two-dimensional Potts models \cite{Fan11}. The reason that we call it instability fixed point will become clear below.

In order to see how to reach the imaginary fixed point, we now solve the flow equation~(\ref{rg}). As $K$ remains a constant, Eq.~(\ref{v}) leads to
\begin{equation}
v^2(l)=\frac{v_0^2A\epsilon}{(A\epsilon+A_6v_0^2){\rm e}^{-\epsilon l}-A_6v_0^2},\label{vl}
\end{equation}
where $v_0$ is the initial coupling constant at $l=0$ as can be easily checked. For $\epsilon<0$ or $d>6$, This solution approaches the Gaussian fixed point continuously as $l\rightarrow\infty$. For $\epsilon>0$, $v^2$ also approaches $v^{*2}$ correctly in the same limit. However, the course of the approach varies. Whereas for a complex $v_0$, the fixed point is continuously reached, for a real $v_0$, $v^2$ diverges at a characteristic scale satisfying $
\exp(-\epsilon l_s)=A_6v_0^2/(A\epsilon+A_6v_0^2)$,
or, to $O(1)$,
\begin{equation}
l_s=\frac{A}{A_6v_0^2}=\frac{AK^3}{K_d\Lambda^{d-6}v_0^2},\label{ls}
\end{equation}
where we have kept $\Lambda$ for dimensional reason, since $l_s$ is dimensionless. Larger than this scale, $v$ becomes purely imaginary and can thus reach the fixed point. From Eq.~(\ref{vl}), the solution to Eq.~(\ref{r}) can be expressed as
\begin{subequations}\label{tla}
\begin{eqnarray}
\tau(l)=\tau_0{\rm e}^{2l}
+\frac{AA_4}{2A_6}\left[{\rm e}^{-\epsilon( l_s-l)}\Phi\left({\rm e}^{-\epsilon( l_s-l)},1,1-\frac{2}{\epsilon}\right)\right.\nonumber\\
\left. -\mathrm{e}^{2l-\epsilon l_s}\Phi\left({\rm e}^{-\epsilon l_s},1,1-\frac{2}{\epsilon}\right)\right]\quad\label{tll}
\end{eqnarray}
for $l\leq l_s$ and
\begin{eqnarray}
\tau(l)={\rm e}^{2l}\left\{\tau_0
-\frac{AA_4}{2A_6}\left[\Phi\left({\rm e}^{-\epsilon l_s},1,1-\frac{2}{\epsilon}\right){\rm e}^{-\epsilon l_s}+{\rm e}^{-2l_s}\right.\right.\nonumber\\
\left.\left.\times\pi\cot\left(\frac{2\pi}{\epsilon}\right)\right]\right\}+ \frac{AA_4}{2A_6} \Phi\left({\rm e}^{-\epsilon( l-l_s)},1,-\frac{2}{\epsilon}\right)\nonumber\\\label{tl}
\end{eqnarray}
\end{subequations}
for $l>l_s$, where $\tau_0\equiv\tau(l=0)$ and $\Phi(z,k,a)=\sum_{n=0}^{\infty}z^n/(n+a)^k$ is the Hurwitz-Lerch transcendent, which has singularity when $a$ equals negative integers and branch cut discontinuity in the complex $z$ plane running from $+1$ to $\infty$. So $\tau$ diverges at $l_s$ too. Because $\Phi(0,1,a)=1/a$, $\tau(l)$ approaches its fixed point value~(\ref{rs}) correctly as $l\rightarrow\infty$. However, to reach this, $\tau_0$ has to assume a particular value that nullifies the terms in the braces in Eq.~(\ref{tl}), i.e., to cancel roughly the accrual up to $l_s$. This may appear strange as $\tau_0$ may be supposed to be zero to suppress the growth of $\tau$ as it is relevant. However, it can be shown that in the $\phi^4$ model for critical phenomena, $\tau_0$ has also to take a particular value to cancel the value at $l=0$ of a function similar to the last one in Eq.~(\ref{tl}).

Our primary task is then to show what the characteristic scale is.
To this end, we employ the theory of nucleation near the spinodal point \cite{Klein83,Unger84}, which is the extension of the nucleation theory near the coexistence curve \cite{Langer67} to the spinodal point in the case of long-range interactions. The only difference is that we consider $\tau$ instead of $h$ studied there. According to these theory \cite{Langer67,Klein83,Unger84,Cahn}, the critically nucleating profile is given by the saddle-point solution of the Euler-Lagrange equation
\begin{equation}
\nabla^2\varphi_s=\tau\varphi_s+\frac{1}{2}\varphi_s^2\label{spe}
\end{equation}
for ${\cal H}$. Upon assuming that the profile depends only on the radius denoted by $x$ and neglecting the first derivative \cite{Klein83,Unger84}, Eq.~(\ref{spe}) becomes
\begin{equation}
\frac{d^2\varphi_s}{dx^2}= \tau\varphi_s+\frac{1}{2}\varphi_s^2\equiv-\frac{\partial{\cal V}}{\partial\varphi_s}\label{spex}
\end{equation}
with a potential ${\cal V}=-f(\varphi_s)$. Equation~(\ref{spex}) describes a particle of unit mass moving in the potential ${\cal V}$ with $\varphi_s$ and $x$ representing its displacement and time, respectively. Therefore, there is a bounded solution if the particle has an initial total energy equal to ${\cal V}(\tau/v)={\cal V}(-2\tau/v)$ as shown in Fig.~\ref{foptfr}(b). The solution is
\begin{equation}
\varphi_s=-\frac{2\tau}{v}\left[1-\frac{3}{2}{\rm sech}^2\left(\frac{1}{2} \sqrt{\frac{-\tau}{K}}x\right)\right],\label{sol}
\end{equation}
which can be checked to have correct limits. This introduces a correlation length $\xi\sim\sqrt{K/|\tau|}$ that diverges as $\tau\rightarrow0$. However, it does not relate to $l_s$. Note that the neglected first derivative acts as a frictional force and has been shown to change only the shape of the profile \cite{Unger84}. In particular, a bound state exists apparently in this case too and the form of the correlation length should keep up to a constant. To create the critical nucleus, there is a free-energy cost \cite{Langer67,Klein83,Unger84,Cahn}
\begin{eqnarray}
\Delta F&=&F(\varphi_s)-F(M_m)\nonumber\\
&=&\int d^dx\left[\frac{1}{2}\tau\varphi_s^2 +\frac{1}{3!}v\varphi_s^3+
\frac{1}{2}K(\nabla \varphi_s)^2 -\frac{2\tau^3}{3v^2} \right]\nonumber\\
&\sim&\frac{K^{d/2}}{v^2}|\tau|^{3-d/2},\label{dF}
\end{eqnarray}
where uses have been made of Eqs.~(\ref{spex}) and (\ref{sol}) and we have neglected higher-order corrections to $\varphi_s$. Note that $\Delta F\sim\xi^d\Delta f$ reasonably and thus its form as given in Eq.~(\ref{dF}) must be valid even the neglected first derivative is taken into account. Equation~(\ref{dF}) shows that for $d>6$, the free-energy cost for nucleation diverges as the spinodal point is approached; while for $d<6$, $\Delta F$ vanishes with $\tau$. This has been argued against existence of the spinodal \cite{Klein83,Unger84} but serves as a physical origin of the borderline critical dimension for the $\epsilon$ expansion.

Now it can be seen that $\Delta F$ is related to the characteristic scale $l_s$.
Apparently, one sees that in $d=6$ on which the $\epsilon$ expansion is based, the $K^3/v^2$ dependence of the free-energy cost $\Delta F$ just coincides with $l_s$. Note that except for possible dimensionless multipliers it is in fact the only combination that can have the dimension of $\Lambda^{d-6}$. Indeed, if one assumes that the dimension of $\varphi$ is $\zeta$, denoted as $[\varphi]=\zeta$, i.e., $\varphi\sim \Lambda^{\zeta}$, then $[\tau]=d-2\zeta$, $[v]=d-3\zeta$, and $[K]=d-2-2\zeta$, because ${\cal H}$ and $F$ are dimensionless as we have absorbed the thermal energy into their definitions. So, $[K^3/v^2]=d-6$.

More explicitly, we can show formally that $l_s$ is just the scale at which $\Delta F$ vanishes. Indeed, using Eq.~(\ref{rg}), we find
\begin{eqnarray}
\frac{d}{dl}\left(\frac{\tau^{\epsilon/2}}{v^2}\right)&=&-\frac{A_6}{A}\tau^{\frac{\epsilon}{2}}-\frac{1} {4}\epsilon A_4\tau^{\frac{\epsilon}{2}-1}+\frac{4-d}{4d}\epsilon A_6\tau^{\frac{\epsilon}{2}}\nonumber\\
&=&-\frac{A_6}{A}\left[1+O(\epsilon)\right],\label{dfl}
\end{eqnarray}
where we have kept $\epsilon$ on the left hand side for definiteness but expanded it on the right hand side. Equation~(\ref{dfl}) indicates that $\tau^{\epsilon/2}/v^{2}$ reduces constantly upon coarse graining. Accordingly, we have
\begin{equation}
\frac{\tau^{\epsilon/2}}{A_6v^2}=\frac{\tau_0^{\epsilon/2}}{A_6v_0^2}-\frac{l}{A}.\label{fl}
\end{equation}
Therefore, to $O(1)$, exactly at $l_s$ given by Eq.~(\ref{ls}), $\tau^{\epsilon/2}/A_6v^{2}$ vanishes. One can check that a constant multiplier does not of course change the result and thus $\Delta F$ indeed vanishes at $l_s$.

Equation~(\ref{fl}) shows, in fact, the effect of $v^{-2}$ and hence the effect of the divergence. Note that Eq.~(\ref{dfl}) is only formal since the neglected terms diverge at $l_s$. However, its right hand side only contains $\tau$, while the flow equation for $\tau^{\epsilon/2}$ itself, for example, contains, besides those $\tau$s in Eq.~(\ref{dfl}), also $v^2$, which is the source of the divergence. Also, the flow equation for $\tau/v$ is not simple either. This seems to indicate that Eq.~(\ref{dfl}) is reasonable. Of course, if we ignore simply $\tau^{\epsilon/2}$ with the $\epsilon$ expansion, Eq.~(\ref{fl}) is just the leading-order terms of $v^{-2}$ as can be checked from Eq.~(\ref{vl}). So, at least to the leading order in the $\epsilon$ expansion, $\Delta F$ indeed vanishes at $l_s$. In fact, since $v^{-2}$ equals zero exactly at $l_s$, all other quantities multiplying $v^{-k}$ with $k>0$ vanish at $l_s$ even though they themselves diverge there. Accordingly, at $l_s$, not only $\Delta F$, but also $\tau/v$ vanish as can be checked numerically from the solutions~(\ref{vl}) and (\ref{tla}). Therefore, at $l=l_s$, the shape of the unform part of the Hamiltonian becomes similar to the chain line with $\tau=0$ in Fig.~\ref{foptfr}(a). Moreover, since $v$ diverges, it becomes extremely steep. In other words, due to the divergence, the system becomes extremely unstable at $l_s$.

Knowing what $l_s$ and the effects of the divergence are, we can then understand their related physics and why the flow becomes imaginary. Given a $v_0$, at least near $d=6$, the $l_s$ divides all the modes into two sets. Whereas the short-wavelength modes that have their wavelengths shorter than that which corresponds to $l_s$ have a finite free-energy cost $\Delta F$ for nucleation, the long-wavelength ones have none. As the RG $\epsilon$ expansion is about the mean-field theory in which no fluctuations exist, a finite $\Delta F$ thus prevents such modes from escaping the free-energy well. Accordingly, the short-wavelength modes probe basically the local metastable state in Fig.~\ref{foptfr}(a) and do not know the existence of the other stable, albeit unbounded, state due to their short-range fluctuations. As a result, they cannot trigger the FOPT and are thus irrelevant to it. On the other hand, the other modes are extremely unstable as they lie just on the true instability point and are clearly relevant to the FOPT.
According to the theories of nucleation \cite{Langer67,Klein83,Unger84}, the free energy of a metastable state becomes complex when the saddle point contributes, because the latter is unstable to perturbation towards the stable state and thus analytical continuation is needed in obtaining $Z$. Thus, although the short-wavelength degrees of freedom can be safely integrated away, the contour for integrating the unstable long-wavelength ones has to be deformed analytically to obtain a meaningful convergent free energy. Indeed, such an integration in Eq.~(\ref{z}) diverges for $\varphi<0$. It has accordingly to be performed along a contour such that the real part of $\varphi^3$ is positive. This can be done by integrating along the imaginary axis for $\varphi<0$, which amounts to replace $\varphi$ with $i\varphi$ and thus $v$ becomes $iv$ along the new contour as speculated~\cite{zhong11}. Therefore, the RG flow of real values corresponds to elimination of the irrelevant degrees of freedom in order to place the system at the true instability point, which exhibits divergence; while the flow beyond becomes imaginary in order to be physical counter-intuitively and drives the system to the fixed point. Note that as the division of modes is set by $\Delta F$, we may well reverse the above derivation: It is physically more plausible that the vanishing of $\Delta F$ at $l_s$ results in a divergent $v^2$ from Eq.~(\ref{fl}).

The above idea can be put in an explicit way. Taking into account the convergence of the $\varphi^3$ model and the division of different degrees of freedom, we may write from the beginning the $\varphi^3$ interaction of ${\cal H}$ as
\begin{equation}
{\cal H}_{\rm in}[\varphi] = \int_{<} d{\bf x}\left(\frac{1}{3!}v_1\varphi^3\right) +\int_{>} d{\bf x}\left(\frac{1}{3!}iv_2\varphi^3\right)\label{h3p}
\end{equation}
with two real parameters $v_1$ and $v_2$, where the two integrations are over spatial extents less and greater, respectively, than that which corresponds to the scale $l_s$. In this parametrization, the first term is irrelevant and disappears beyond $l_s$ upon coarse graining and can thus be ignored; while the second one is the only relevant interaction for a meaningful convergent description of the real FOPT from the instability point to the left side in Fig.~\ref{foptfr}(a). The relevant model coincides then with the one proposed for the Yang--Lee edge singularity~\cite{Fisher78}. Its RG behaves benign as it has no divergences and a real fixed point that is just the imaginary one found above. This therefore confirms the perturbation analysis and the physics of the imaginary fixed point. Moreover, the model justifies directly without the mapping employed previously~\cite{zhong11} that the edge singularity and the FOPTs fall in the same universality class as they now share the same relevant model. However, the real Hamiltonian~(\ref{h3}) itself can automatically find its proper fixed point through divergence. Note that it appears that nucleation may also be governed by the same fixed point.

Concluding, we have shown that, at the characteristic scale at which the RG flow diverges, the free-energy cost for nucleation out of the metastable state vanishes and the system becomes extremely unstable and then enters an imaginary domain for convergence.
One may still seek how the perturbation expansion in $d=6$ in which $\Delta F$ does not depend on $\tau$ can be carried over to other dimensions. However, the clear physical picture emerges should suffice to show the imaginary fixed point of the $\varphi^3$ model does be reachable physically and control the scaling and universality of driven FOPTs.

I thank Shuai Yin, Yantao Li, Guangyao Li, and Ning Liang for their useful discussions. This work was supported by NNSFC (Grant No. 10625420) and FRFCUC.

\end{document}